\documentclass[conference]{IEEEtran}
\IEEEoverridecommandlockouts
\usepackage{cite}
\usepackage{amsmath,amssymb,amsfonts}
\usepackage{algorithmic}
\usepackage{graphicx}
\usepackage{textcomp}
\usepackage{xcolor}
\usepackage{siunitx}
\usepackage{multirow}
\def\BibTeX{{\rm B\kern-.05em{\sc i\kern-.025em b}\kern-.08em
    T\kern-.1667em\lower.7ex\hbox{E}\kern-.125emX}}
\begin{document}

\title{X-pSRAM: A Photonic SRAM with Embedded XOR Logic for Ultra-Fast In-Memory Computing
}

\author{\IEEEauthorblockN{Md Abdullah-Al Kaiser$^*$, Sugeet Sunder$^\dagger$, Ajey P. Jacob$^\dagger$, and Akhilesh R. Jaiswal$^*$}
\IEEEauthorblockA{\textit{$^*$University of Wisconsin--Madison, WI, USA, $^\dagger$USC Information Sciences Institute, CA, USA} \\
mkaiser8@wisc.edu, sunder@isi.edu, ajey@isi.edu, akhilesh.jaiswal@wisc.edu}
}

\maketitle

\begin{abstract}
Traditional von Neumann architectures suffer from fundamental bottlenecks due to continuous data movement between memory and processing units, a challenge that worsens with technology scaling as electrical interconnect delays become more significant. These limitations impede the performance and energy efficiency required for modern data-intensive applications. In contrast, photonic in-memory computing presents a promising alternative by harnessing the advantages of light, enabling ultra-fast data propagation without length-dependent impedance, thereby significantly reducing computational latency and energy consumption. This work proposes a novel differential photonic static random access memory (pSRAM) bitcell that facilitates electro-optic data storage while enabling ultra-fast in-memory Boolean XOR computation. By employing cross-coupled microring resonators and differential photodiodes, the XOR-augmented pSRAM (X-pSRAM) bitcell achieves at least 10 GHz read, write, and compute operations entirely in the optical domain. Additionally, wavelength-division multiplexing (WDM) enables n-bit XOR computation in a single-shot operation, supporting massively parallel processing and enhanced computational efficiency. Validated on GlobalFoundries' 45SPCLO node, the X-pSRAM consumed 13.2 fJ energy per bit for XOR computation, representing a significant advancement toward next-generation optical computing with applications in cryptography, hyperdimensional computing, and neural networks.
\end{abstract}


\section{Introduction}
The increasing demand for data-intensive applications, such as artificial intelligence, machine learning, and large-scale simulations, necessitates ultra-fast and efficient computation. However, traditional von Neumann architectures suffer from a fundamental limitation: the separation of memory and processing units. This segmentation results in reduced speed and energy efficiency due to the frequent data transfers between memory and the processor \cite{memory_wall_bottleneck}. In-memory computing (IMC) addresses this issue by enabling computations directly within memory, minimizing data movement and enhancing throughput \cite{imc_ref1, imc_ref2}. Nevertheless, interconnect scaling introduces increased resistance and capacitance, leading to higher delays and energy consumption, which further limits the efficiency of conventional electrical systems \cite{interconnect_prb1, interconnect_prb2}. In contrast, optical computing presents a promising alternative by utilizing waveguides for high-speed, low-loss data transmission, effectively overcoming the limitations of electrical interconnects \cite{optical_comp, optics_advantage1, optics_advantage2}. Moreover, wavelength-division multiplexing (WDM)-based approaches allow for massively parallel optical computations, significantly enhancing computational efficiency and throughput \cite{multiplex_optical_domain}. Therefore, by integrating photonic computing with in-memory approaches, a scalable and energy-efficient solution can be realized, offering a promising path to overcoming the von Neumann bottleneck and enabling the next generation of high-performance computing.

Among various in-memory compute operations, the XOR (exclusive OR) is a fundamental Boolean operation that plays a critical role in a wide range of applications, including large-scale data verification, encryption, pattern matching, and classification algorithms \cite{xnor_net, xor_net, xnor_bnn, secure_XOR_CIM, xor_in_AI, xnor_sram, xor_verification, x_sram, xor_copy, hd_computing1, hd_computing2}. Consequently, an ultra-fast and energy-efficient in-situ in-memory XOR operation can dramatically enhance energy efficiency across these diverse application domains. Despite this, although several SRAM-based in-memory XOR compute macros have been reported, the latency per compute operation typically ranges from 0.8 to 3 ns \cite{xnor_sram, x_sram}. Additionally, various optical implementations have utilized photonic crystal technology for XOR computations, but these necessitate significant modifications to existing standard CMOS foundry processes \cite{PhC_xor1, PhC_xor2, PhC_xor3, PhC_xor4}. While microring resonator (MRR)-based XOR implementations have been demonstrated, they rely on external control of the MRR via DC pads, which can limit their potential as on-chip macro processing blocks. Furthermore, DC control may restrict the operational speed while consuming substantial modulation power \cite{eo_xor_dc1, eo_xor_dc2, eo_xor_dc3}. However, these photonic XOR implementations in the optical domain only perform the Boolean computation and do not execute in-memory XOR computation.

In this context, we introduce a novel XOR-augmented photonic SRAM (X-pSRAM) bitcell that stores binary data and performs in-situ XOR computation with the incoming data. Our design utilizes CMOS-compatible, fabrication-friendly photonic components, such as microring resonator (MRR), photodiode (PD), multi-mode interference coupler (MMI), and power splitter (PS). By employing wavelength-division multiplexing (WDM), we enable highly parallel n-bit XOR operations within the memory array in a single-shot execution. To the best of our knowledge, this is the first photonic compute block that simultaneously enables both data storage and in-memory XOR computation using CMOS-compatible, fabrication-friendly silicon photonic elements. The key contributions of this paper are as follows:

\begin{figure*}[!t]
\centering
\includegraphics[width=1\linewidth]{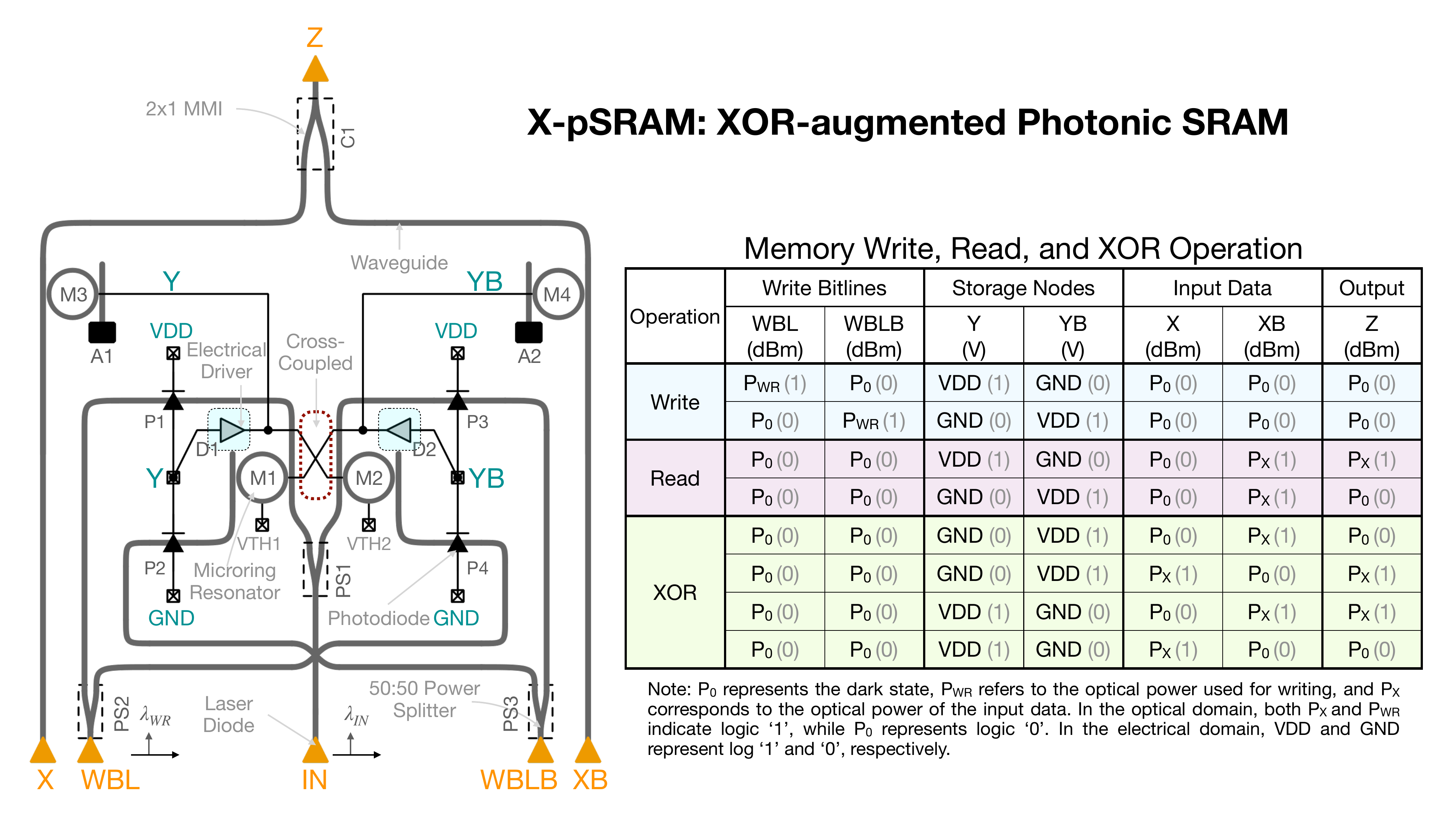}
\caption{Schematic and operational overview of the XOR-augmented photonic SRAM (X-pSRAM) bitcell.}
\label{fig_xor}
\end{figure*}

\begin{enumerate} 
    \item We propose X-pSRAM, a photonic memory bitcell that enables in-situ XOR computation between the stored and input data using CMOS-compatible photonic components. 
    \item We leverage WDM to achieve massively parallel n-bit XOR computation in a single-shot operation. 
    \item We validate the X-pSRAM design on GlobalFoundries' 45SPCLO node, demonstrating 13.2 fJ/XOR at 10 GHz.
\end{enumerate}

\section{XOR-augmented pSRAM Bitcell (X-pSRAM)} \label{sec_xor}
Figure \ref{fig_xor} illustrates the schematic and operational description of the XOR-augmented photonic SRAM (X-pSRAM) bitcell, which is constructed using CMOS-compatible, fabrication-friendly silicon photonic components. The bitcell includes optical waveguides, microring resonators (MRRs), photodiodes (PDs), optical power splitters (PSs), a multi-mode interference coupler (MMI), and passive optical absorbers (A). The waveguides confine and guide optical light along a defined path, while the MRR functions as an optical switch, consisting of a ring waveguide and two bus waveguides. When the MRR is in resonance, light is coupled into the ring waveguide and exits through the drop port; when out of resonance, the thru port receives the majority of the incoming light \cite{mrm_ref1, mrm_ref2}. The photodiode converts incident light into electrical current, the power splitter evenly distributes the input power to two output ports, and the MMI coupler combines multiple optical inputs into a single output. The passive absorbers reduce internal reflections by absorbing excess light within the chip. The following sections provide a detailed explanation of the X-pSRAM structure, including its hold, write, read, and XOR-compute operations.

\subsection{Structure of the X-pSRAM}
In the schematic, M1–M4 denote the microring resonators, P1–P4 correspond to photodiodes, PS1–PS3 represent 50:50 power splitters, C1 denotes the multi-mode interference coupler, and A1–A2 function as passive optical absorbers. The nodes positioned between photodiodes P1 and P2 (similarly, P3 and P4) are identified as \si{Y} and \si{YB}, serving as the electro-optic storage elements of the X-pSRAM bitcell \cite{pSRAM_design}. Node \si{Y} (\si{YB}) modulates M2 (M1) in a cross-coupled configuration through electrical driver D1 (D2), forming a bistable latch. The thru and drop bus waveguides of M1 (M2) are connected to the waveguides of the photodiodes P1 (P3) and P2 (P4), respectively. This configuration enables reliable binary data storage, where \si{Y} holds the primary data state and \si{YB} maintains its complementary counterpart. An optical bias laser (\si{\lambda_{IN}}) is connected to the input 50:50 power splitter PS1, which directs half of the laser power to the input bus waveguide of the latch MRRs (M1 and M2). The wavelength of the laser source (\si{\lambda_{IN}}) is selected to resonate with the MRRs when a voltage of VDD is applied across them. VTH1 and VTH2 are the thermal control ports of latch MRRs M1 and M2, enabling post-fabrication calibration to correct any resonance mismatches.

The write bitline waveguides, WBL (WBLB), are connected to the photodiodes P1 and P4 (P2 and P3) via power splitters PS2 (PS3), enabling data flipping within the storage nodes \si{Y} and \si{YB}. Additionally, the read/input bitline waveguides connected to \si{X} (\si{XB}) are used for performing single-ended read operations of the stored data or for executing Boolean XOR operations between the input data (\si{X}) and the stored data (\si{Y}). These read/input waveguides are linked to the input ports of the microring resonators M3 and M4. The source wavelength of the input/read bitline waveguides is set to match \si{\lambda_{IN}}, ensuring that when a voltage of \si{VDD} is applied across M3 and M4, they will resonate. The drop-ports of M3 and M4 are connected to passive absorbers A1 and A2 to absorb light resonating with M3 and M4, preventing any potential reflections. The thru-ports of M3 and M4 are combined using a MMI coupler C1, generating the output \si{Z}, which serves as the bitline output for either the read or in-memory XOR compute operation.

\subsection{Hold Operation}
The storage node of the X-pSRAM bitcell, as shown in Fig. \ref{fig_xor}, can retain data (\si{VDD} for logic 1, GND for logic 0) as long as both the optical bias (input laser source at the IN port in Fig. \ref{fig_xor}) and the electrical bias (\si{VDD}) are maintained. For example, when \si{Y} = 1 (\si{VDD}) and \si{YB} = 0 (GND) are stored, the electrical driver D1 drives M2 to resonate with the incoming light from the IN port. As a result, the drop port of M2 receives more light than the thru port, causing photodiode P4 to generate more current (providing a lower resistance path to GND) compared to P3. This causes the \si{YB} node to be pulled down to GND due to the resonance condition of M2. Consequently, the \si{YB} node drives MRR M1 through electrical driver D2, moving it out of resonance with the input light. As a result, the input light from the IN port is directed to photodiode P1 via the thru port of M1, while P2 receives minimal light since M1 is out of resonance. This leads to node \si{Y} being pulled high through the low-resistance path created by P1 to \si{VDD}. This cross-coupled back-to-back driving mechanism ensures stable data retention in the storage nodes of the X-pSRAM bitcell. A similar process holds for the opposite data state, where \si{Y} = 0 and \si{YB} = 1, causing M1 to be in resonance and M2 to be out of resonance. 

\subsection{Write Operation} \label{sec_xpsram_write}
The operational overview shown in Fig. \ref{fig_xor} describes the process of writing data to the storage nodes of the X-pSRAM bitcell. During a write operation, the input/read waveguides (\si{X} and XB) remain in a dark state (\si{P_0}), and the combined output waveguide \si{Z} also outputs no light, as both \si{X} and \si{XB} are in the dark. To initiate the write operation, a write laser source sends light through the write bitline waveguide in a differential manner. It is important to note that the write laser power (\si{P_{WR}}) must be higher than the laser power connected at the IN port to successfully flip the data. For example, assuming the X-pSRAM currently stores data \si{Y} = 0 (GND) and \si{YB} = 1 (\si{VDD}), and the goal is to write \si{Y} = 1 (\si{VDD}) and \si{YB} = 0 (GND), the write laser source transmits power through WBL (\si{P_{WR}}), while WBLB (\si{P_0}) remains in the dark state. Consequently, photodiodes P1 and P4 receive more optical power via the write bitline waveguides, causing \si{Y} to be pulled high to \si{VDD} through a low-resistance path via P1, and \si{YB} to be pulled low to GND through a low-resistance path via P4. As a result, \si{Y} enables resonance with the input light through M2, while \si{YB} drives M1 out of resonance. The new data is then stably stored using the cross-coupled configuration. The opposite data (\si{Y} = 0, \si{YB} = 1) can also be written by supplying more light through WBLB (\si{P_{WR}}) while keeping WBL (\si{P_0}) in the dark state.

\subsection{Read Operation} \label{sec_xpsram_read}
The X-pSRAM bitcell can perform the read operation of the stored data (\si{Y}) through the read bitline waveguides, either via \si{X} or XB. The thru ports of the read/compute MRRs M3 and M4 are connected to the output waveguide \si{Z}. To enable an active high output logic during the read operation, we utilize the \si{XB} port. During the read process, the waveguides connected to \si{X} remain in the dark state (\si{P_0}), and optical power is sent through the \si{XB}-connected waveguide (\si{P_X}). \si{XB} is linked to the read ring M4, which is driven by the complementary data \si{YB}. If the stored data in \si{Y} is 1 (and \si{YB} is GND), M4 will be out of resonance, allowing \si{Z} to receive the incoming light via \si{XB}. However, if the stored data in \si{Y} is 0 (and \si{YB} is \si{VDD}), M4 will resonate with the incoming light through \si{XB}, causing no light to be output through \si{Z}. The resonating light will then be absorbed by absorber A2. In summary, when the X-pSRAM stores a 1, optical power will be received by \si{Z}; when it stores a 0, \si{Z}  will remain in the dark state. Note, an active-low read operation can also be performed utilizing the \si{X} port during the read operation. 

\subsection{XOR Operation} \label{sec_xpsram_xor}
The X-pSRAM bitcell can execute an in-situ XOR operation between the input data (\si{X}) and the stored data (\si{Y}). During this XOR-compute operation, the write bitline waveguides remain in a dark state (\si{P_0}). Note, optical power \si{P_X} denotes logic-1, and the dark state (\si{P_0}) represents logic-0. The input data is transmitted through read/compute waveguides connected to the \si{X} and \si{XB} ports, while the M3 and M4 microring resonators (MRRs) perform the XOR computation. When both the input data (\si{X} = \si{P_0}, \si{XB} = \si{P_X}) and the stored data (\si{Y} = GND, \si{YB} = \si{VDD}) are logic 0, the output node \si{Z} will receive no light and remain in the dark state (\si{P_0}). Since the input data is logic 0, no light is received at the \si{X} port, and the light from the \si{XB} port will resonate with MRR M4 because \si{YB} = \si{VDD}, resulting in no light passing through the MMI coupler C1. Consequently, \si{Z} remains in the dark state. Similarly, when the input data (\si{X} = \si{P_X}, \si{XB} = \si{P_0}) and the stored data (\si{Y} = \si{VDD}, \si{YB} = GND) are logic 1, the light entering the \si{X} port will resonate through MRR M3 and be absorbed by A1. This prevents any light from reaching the input ports of the C1, causing \si{Z} to remain in the dark state (\si{P_0}).

In contrast, when the input data (\si{X} = \si{P_X}, \si{XB} = \si{P_0}) is logic 1 and the stored data (\si{Y} = GND, \si{YB} = \si{VDD}) is logic 0, MRR M3 will be out of resonance. As a result, the input light from the \si{X} port will pass through the C1, leading to a high optical power output at the \si{Z} node. Similarly, in the opposite scenario when the input data (\si{X} = \si{P_0}, \si{XB} = \si{P_X}) is logic 0 and stored data is (\si{Y} = \si{VDD}, \si{YB} = GND) is logic 1, the \si{YB} node keeps MRR M4 out of resonance. This allows light from the \si{XB} port to be transmitted to the C1, which in turn results in a higher optical power output at the \si{Z}  node. In summary, the X-pSRAM performs an in-situ XOR operation: when the input and stored data are the same, no light is output, while when the input and stored data differ, high optical power is output at the \si{Z} node.

\section{X-pSRAM Array Architecture} \label{sec_xor_array}
\begin{figure}[!t]
\centering
\includegraphics[width=1\linewidth]{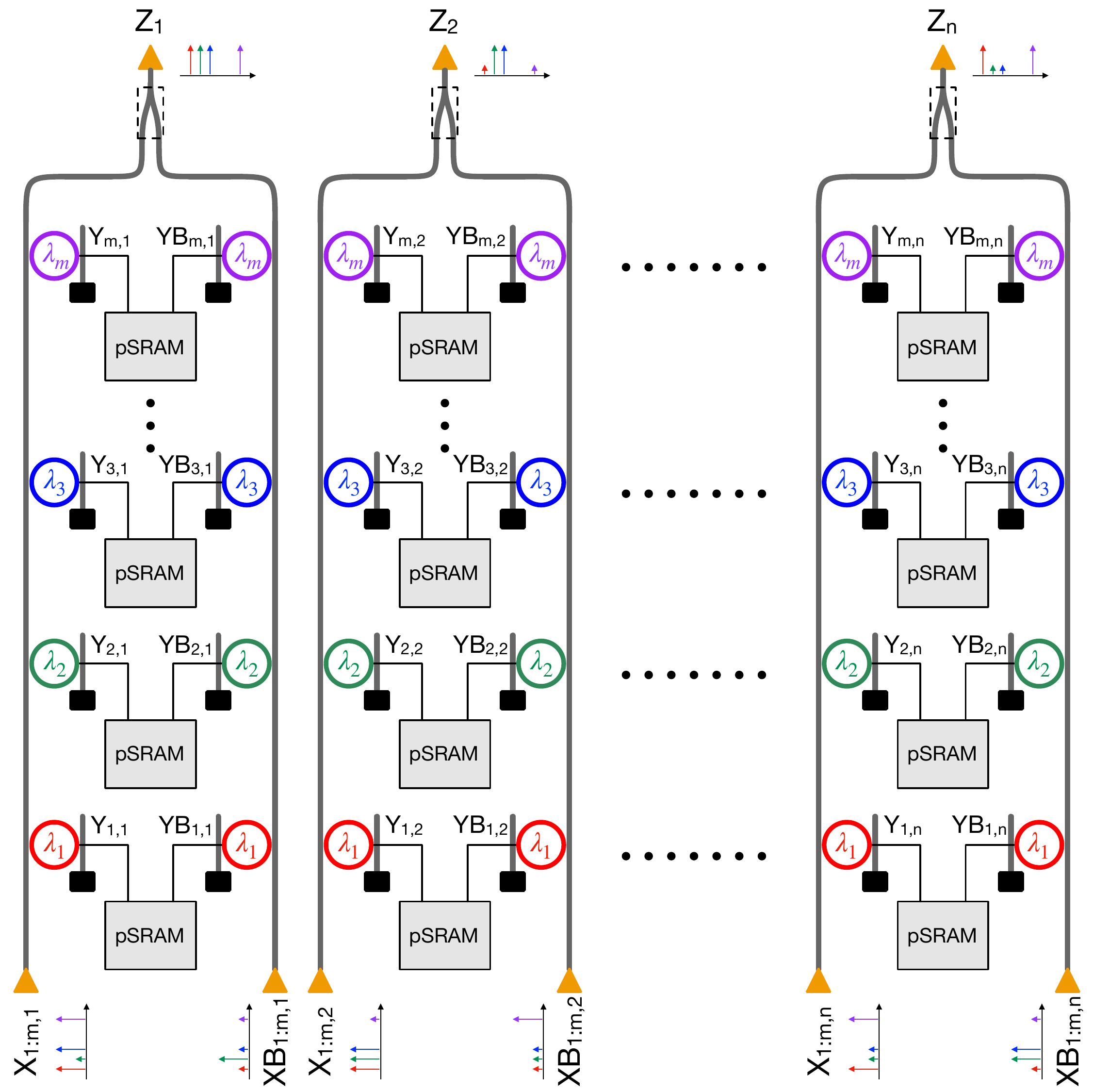}
\caption{X-pSRAM Array Architecture for n-bit XOR Computation leveraging Wavelength-Division Multiplexing.}
\label{fig_xor_array}
\end{figure}

Fig. \ref{fig_xor_array} illustrates the architecture of the X-pSRAM array, which enables the computation of n-bit XOR operations by leveraging wavelength-division multiplexing (WDM) in optical computing. The array consists of an \si{m \times n} structure, where \si{m} represents the number of rows and \si{n} the number of columns.
The X-pSRAM schematic, as shown in Fig. \ref{fig_xor}, is divided into two components: the gray square boxes represent the photonic memory structure, while the compute microring resonators (M3 and M4) are also depicted. The input waveguides, shared across the rows, are denoted as \si{X_{1:m,1}}, \si{X_{1:m,2}}, \dots, \si{X_{1:m,n}}, while the output ports are represented by \si{Z_1}, \si{Z_2}, \dots, \si{Z_n}.

The compute microring resonators of each row can resonate at a specific wavelength, which can be modulated by the geometry (length or radius) of the MRR. For example, the compute microring resonators in row-1, row-2, and row-\si{m} resonate at wavelengths \si{\lambda_1}, \si{\lambda_2}, \dots, and \si{\lambda_m}, respectively. To enable the n-bit in-situ XOR operation, WDM is employed for input data, where each input bit is also encoded at a distinct wavelength. Specifically, input data-1 is encoded in \si{\lambda_1}, input data-2 in \si{\lambda_2}, and so on, for each of the \si{Xs} and \si{XBs} waveguides.

Since each compute microring resonator (MRR) is designed to resonate at a specific wavelength, input data-1, encoded in \si{\lambda_1}, will only interact with the stored data in row-1, as the resonators in row-1 are precisely tuned to \si{\lambda_1}. This wavelength-specific resonance ensures that each row’s compute microring resonator processes only the data corresponding to its designated wavelength. The in-situ XOR computation is performed by the MRRs, and the XOR output is obtained from the thru port of each resonator. The responses from all rows are then combined through the common MMI coupler per column, which aggregates the results. The XOR output is made available at the \si{Z} node via wavelength-division multiplexing (WDM). For example, \si{\lambda_1} at the \si{Z} node corresponds to the XOR output between input data-1 and the stored data in row-1, \si{\lambda_2} represents the XOR output between input data-2 and stored data in row-2, and so on for the other rows. This approach allows for parallel processing of multiple XOR operations, with each wavelength carrying the result for a specific row, facilitating efficient and scalable computation.

Using WDM, the proposed X-pSRAM can perform massively parallel n-bit XOR operations. Additionally, the array structure can also be used to add the XOR outputs by placing photodiodes at the \si{Z} node, enabling it to perform kernel multiplication and addition in binary neural networks. Furthermore, cosine similarity can be estimated by analyzing the optical power output from the \si{Z} node. 

\section{Simulation Results} \label{sec_results}
This section presents the verification results and performance metrics of the proposed X-pSRAM bitcell and X-pSRAM array, simulated using the monolithic GF45SPCLO technology node.

\subsection{X-pSRAM Write, Hold, and Read Operation}

\begin{figure}[!t]
\centering
\includegraphics[width=1\linewidth]{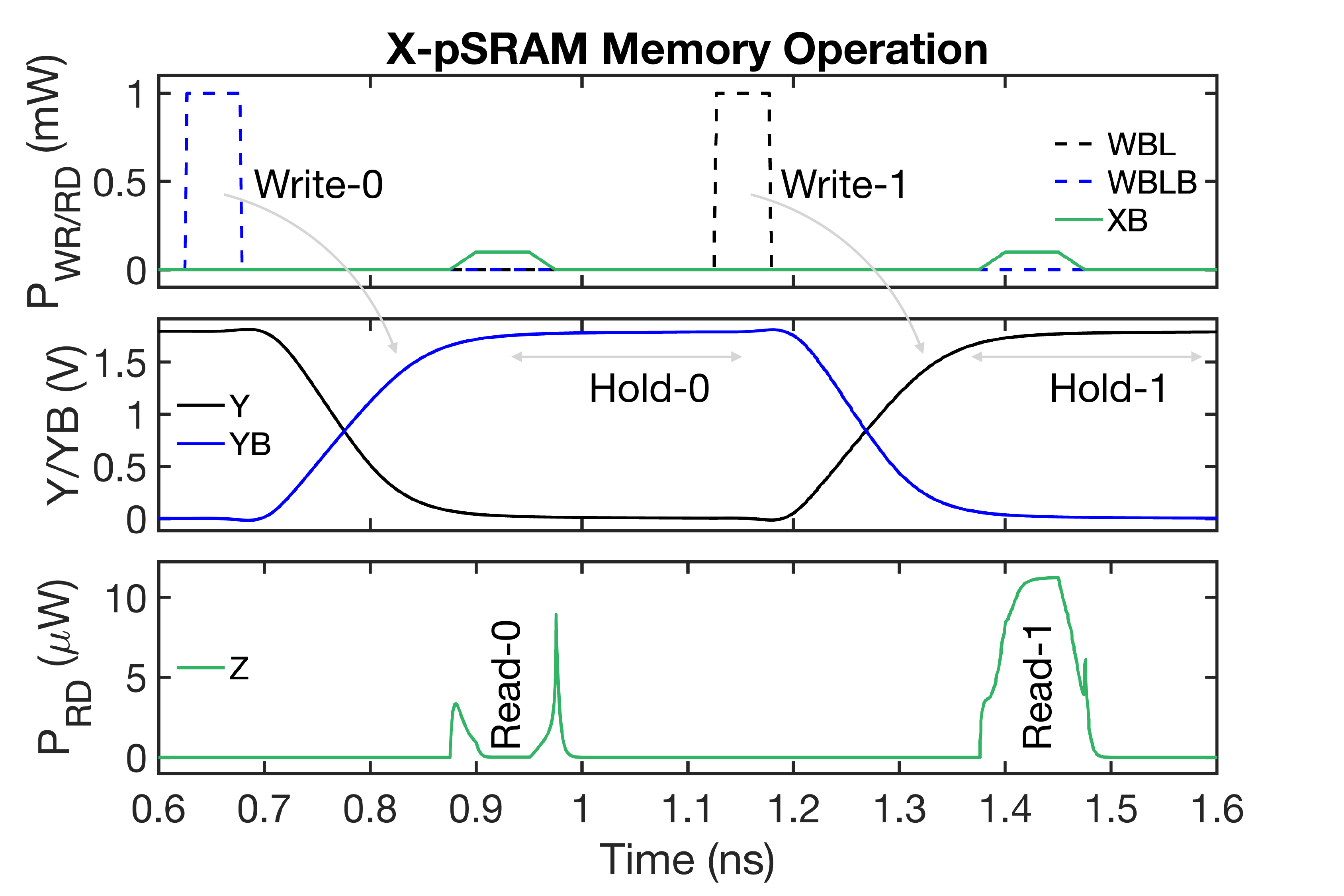}
\caption{Verification of the write, hold and read operations in X-pSRAM.}
\label{fig_sim_memory}
\end{figure}

Figure \ref{fig_sim_memory} illustrates the write, hold, and read verification of the proposed X-pSRAM bitcell. The top subplot shows the optical power input applied through the WBL, WBLB, and \si{XB} ports during write and read operations. The middle subplot represents the state of the electro-optic storage nodes \si{Y} and \si{YB}, while the bottom subplot depicts the optical power output at the \si{Z} port.

The write operation (discussed in subsection \ref{sec_xpsram_write}) is initiated by transmitting a 50 ps-wide write pulse with 1 mW power differentially through WBL and WBLB. When WBLB receives the write pulse, \si{Y} is driven to GND, and \si{YB} transitions to \si{VDD} due to the cross-coupled configuration of the MRRs and PDs, occurring at approximately 0.8 ns. Conversely, when the opposite pulse is applied (WBL = 1 mW, 50 ps; WBLB = 0), the stored data is inverted. As seen in the figure, \si{Y} switches from GND to \si{VDD}, while \si{YB} changes from \si{VDD} to GND at around 1.3 ns.

For the read operation (discussed in subsection \ref{sec_xpsram_read}), an optical laser signal is applied at the \si{XB} port, and the output is measured at the \si{Z} port. The bottom subplot indicates that when the stored data is 0 (\si{Y} = GND, \si{YB} = \si{VDD}), the \si{Z} port receives negligible optical power, observable at around 0.93 ns. In contrast, when the stored data is 1 (\si{Y} = \si{VDD}, \si{YB} = GND), the \si{Z} port exhibits a higher optical power output (at around 1.43 ns). By setting a threshold on the optical power at the MMI coupler C1, the output can be seamlessly converted into an electrical signal or directly used as an optical signal for subsequent processing stages. Note that the optical power received at the \si{Z} port accounts for losses due to the read MRR and coupler.

In the simulation, the latch rings M1 and M2 have a radius of 7.5 \si{\micro m}, with bus and drop waveguide gaps of 180 nm and 395 nm, respectively. The laser connected to the IN port is set at 10 \si{\micro W} with a wavelength of 1310.52 nm, which matches the resonance wavelength of M1 and M2. The read rings M3 and M4 have the same radius as M1 and M2 but feature a bus waveguide gap of 200 nm. The simulation results demonstrate that the write operation requires only a 50 ps pulse, enabling a memory update speed of up to 20 GHz for the X-pSRAM bitcell. For read operations, a 100 ps pulse with 100 \si{\micro W} power is used, though higher input power can be applied to enhance the optical output at the \si{Z} port. 


\begin{figure}[!b]
\centering
\includegraphics[width=1\linewidth]{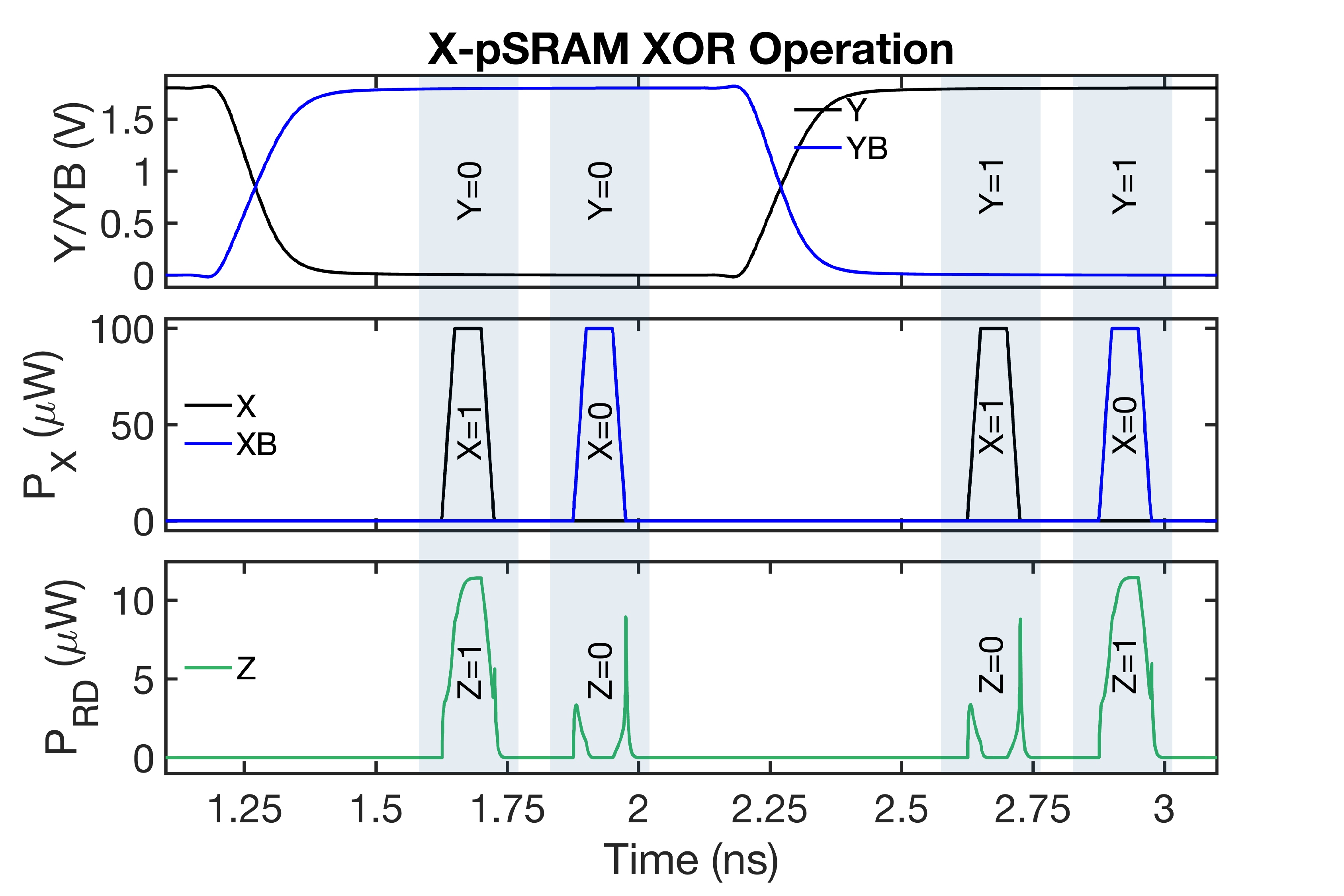}
\caption{Verification of the XOR-compute operation in X-pSRAM.}
\label{fig_sim_xor}
\end{figure}

\subsection{X-pSRAM XOR-compute Operation}
Fig. \ref{fig_sim_xor} illustrates the XOR-compute operation between the stored data (\si{Y}) and the input data (\si{X}). The top subplot represents the stored values at nodes \si{Y} and \si{YB}, the middle subplot shows the optical input data applied through the \si{X} and \si{XB}  ports, and the bottom subplot displays the resulting optical power at the \si{Z} port. To validate all possible logic combinations, we first store a logic 0 in the X-pSRAM bitcell (\si{Y} = GND, \si{YB} = \si{VDD}) and then perform the XOR operation with input values \si{X} = 1 (\si{X} = \si{P_X}, \si{XB}  = \si{P_0}) and \si{X} = 0 (\si{X} = \si{P_0}, \si{XB} = \si{P_X}), respectively. Next, we store logic 1 (\si{Y} = \si{VDD}, \si{YB} = GND) and repeat the XOR computation for input values \si{X} = 1 and \si{X} = 0. 

As shown in the Fig. \ref{fig_sim_xor}, when the stored data is 0 (\si{Y} = GND, \si{YB} = \si{VDD}) and the input data is 1 (\si{X} = \si{P_X}, \si{XB} = \si{P_0}), the \si{Z} port produces a high optical power output at approximately 1.67 ns. Conversely, when the input data is 0 (\si{X} = \si{P_0}, \si{XB} = \si{P_X}), both stored and input data are identical, resulting in a minimal optical power output at \si{Z} (around 1.93 ns). Next, the stored data is updated from 0 to 1 (\si{Y} = \si{VDD}, \si{YB} = GND) via the write bitline waveguides. From the bottom subplot, it can be observed that when the input data is 1 (\si{X} = \si{P_X}, \si{XB} = \si{P_0}), the \si{Z} output remains low (minimal optical power at approximately 2.67 ns). However, when the input data is 0 (\si{X} = \si{P_0}, \si{XB} = \si{P_X}), the \si{Z} output exhibits higher optical power (around 2.93 ns). These results confirm the correct functionality of the XOR logic operation between the stored and input data. A 100 ps pulse with 100 \si{\micro W} optical power has been utilized to perform the per-bit XOR-compute operation.

\begin{figure}[!b]
\centering
\includegraphics[width=1\linewidth]{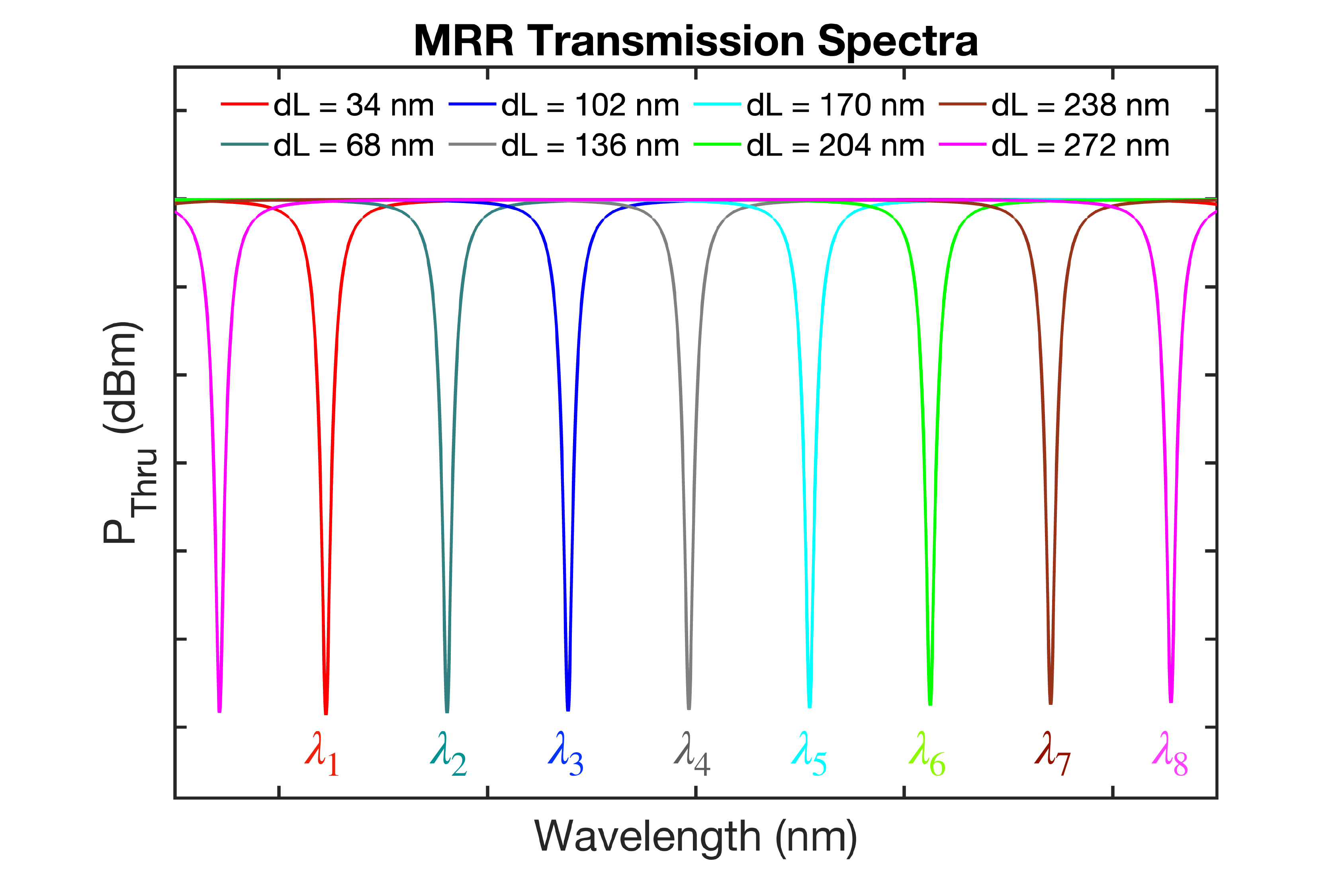}
\caption{Resonance wavelength tuning for WDM-based XOR computation.}
\label{fig_sim_wdm}
\end{figure}

\begin{figure*}[!t]
\centering
\includegraphics[width=1\linewidth]{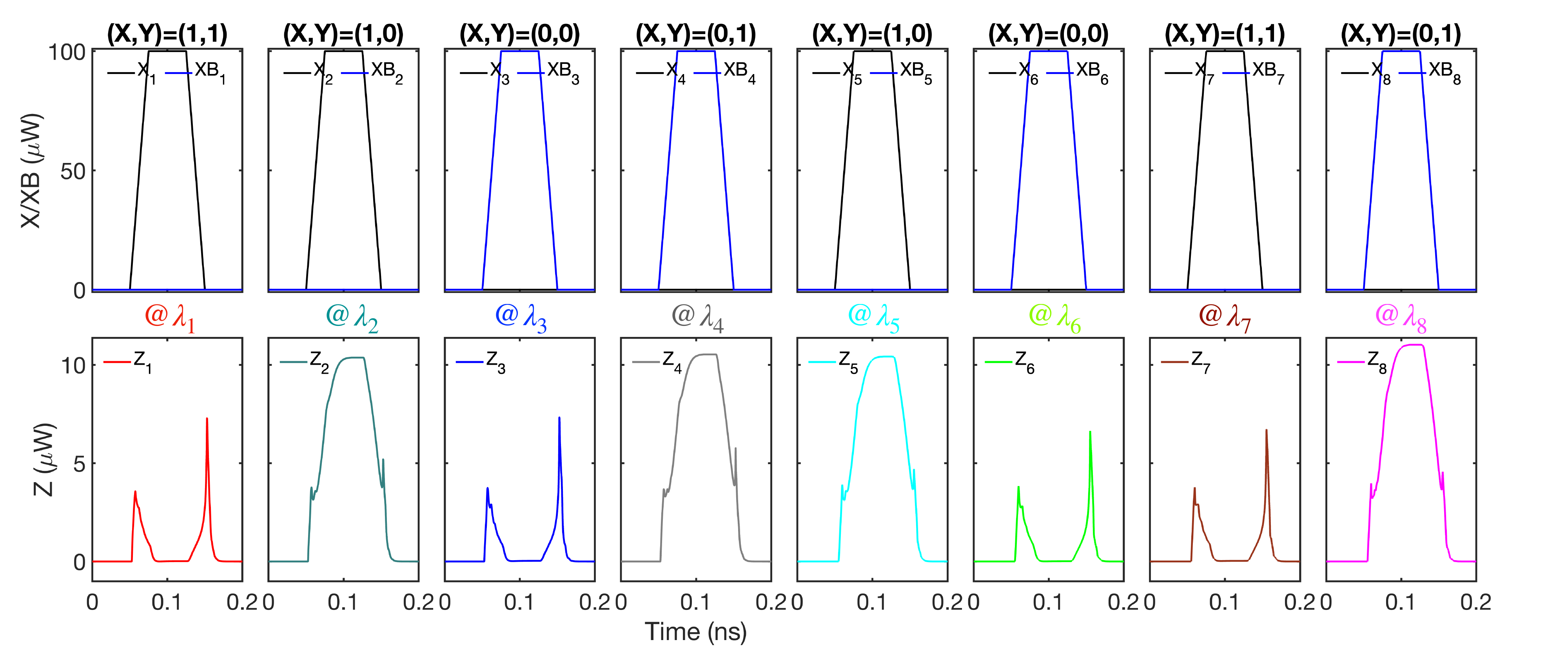}
\caption{Verification of 8-bit parallel XOR-compute operation using wavelength-multiplexed X-pSRAM array.}
\label{fig_sim_array}
\end{figure*}

\begin{table*}[!t]
\caption{Performance comparison of various XOR compute.}
\label{table_comp}
\begin{center}
\begin{tabular}{|c|c|c|c|c|c|} \hline
Reference & Latency  & Energy/bit  & Method & Area & Supported Operations \\
          & (ns)     & (fJ)  &   & (mm\si{^2})  &   \\ \hline
\cite{x_sram} & 3 & 17.25 & 8T-SRAM IMC  & \si{<3\times10^{-6}}$^\#$ & Memory, XOR, NAND, NOR \\
\cite{xnor_sram} & 0.85 & 3679$^*$ & SRAM IMC & \si{<3\times10^{-6}}$^\#$ &  Memory, XNOR   \\
\cite{eo_xor_dc1} & 10 & --- & Boolean Photonic XOR & 0.48 &  XOR \\
\cite{eo_xor_dc3} & 0.05$^\dagger$ & --- & Boolean Photonic XOR & 0.56 &  XOR \\
\cite{optical_FF} & 0.2 & 7960 $^\ddagger$  &  InP Optical Flip-flop & 12 & Memory \\
\textbf{This Work} & 0.1 & 13.15  & Photonic IMC XOR & 0.1 &  Memory, XOR, XNOR  \\ \hline
\multicolumn{4}{l}{$^*$ \scriptsize{Large \si{VDD} is considered to get the best latency number.}} \\
\multicolumn{4}{l}{$^\dagger$ \scriptsize{High-speed power-hungry electro-optic modulator is required.}} \\
\multicolumn{4}{l}{$^\ddagger$ \scriptsize{Power requirement for SET is 16 dBm.}} \\
\multicolumn{4}{l}{$^\#$ \scriptsize{Estimated using 45 nm CMOS Technology.}}
\end{tabular}
\end{center}
\end{table*}

\subsection{n-bit WDM XOR-compute Operation}
To enable WDM-based n-bit parallel XOR-compute operations (as discussed in Section \ref{sec_xor_array}), the geometry of the compute ring must be carefully adjusted to resonate at specific wavelengths. In GlobalFoundries' 45SPLCO PDK, this tuning is achieved through a tunable parameter (dL), which allows precise control over the resonance wavelength of the microring resonators. By incrementally varying the adjustment length (dL) in steps of 34 nm, we can configure the rings to resonate at eight distinct wavelengths within a single free spectral range (FSR), ensuring minimal crosstalk between channels. Fig. \ref{fig_sim_wdm} illustrates the relationship between the resonance wavelength and the adjusted ring length for a ring radius of 7.5 \si{\micro m}. Utilizing this geometry, we implement an 8-way wavelength-division multiplexing (WDM) scheme to enable highly parallel 8-bit XOR computations in a single-shot operation, enhancing computational throughput and efficiency.

Fig. \ref{fig_sim_array} demonstrates the verification of an n-bit (n = 8) parallel XOR-compute operation in a single shot, as discussed in Section \ref{sec_xor_array}. In this verification, the stored data is \si{Y} = \si{10010011}, and the input data is \si{X} = \si{11001010}, with the expected output of the XOR computation being \si{Z} = \si{01011001}. The 8-bit pSRAM stores the input data, and the compute rings for each row are tuned to distinct wavelengths (\si{\lambda_1}, \si{\lambda_2}, \dots, \si{\lambda_8}) as shown in Figure \ref{fig_sim_wdm}. The input data is also encoded at the corresponding wavelengths (\si{\lambda_1}, \si{\lambda_2}, \dots, \si{\lambda_8}), enabling the wavelength-specific XOR computation for each bit position. For example, the leftmost subplots indicate that when the input data is 1 (\si{X_1 = P_X} and encoded in \si{\lambda_1}) and the stored data is 1, the output at the \si{Z} node is 0 (\si{Z_1} has minimal optical power, indicating no power at \si{\lambda_1}). The bottom subplots show the optical power at different wavelengths (\si{Z_1} for \si{\lambda_1}, \si{Z_2} for \si{\lambda_2}, and so on). These results confirm that the X-pSRAM can successfully perform 8-bit XOR operations using 8 different wavelengths in a single shot.


\subsection{Performance Metrics Calculation}
Key performance metrics, including compute latency and speed per XOR operation, have been analyzed for our proposed and validated X-pSRAM bitcell. Each XOR computation utilizes 100 \si{\micro W} input laser power with a 100 ps pulse width. Considering a 10 \si{\micro W} bias laser connected to the IN port, along with electrical energy consumption for photodiode biasing and electrical drivers, the estimated average energy per bit computation is 13.2 fJ (optical: 11 fJ, electrical: 2.2 fJ). The compute latency per operation is 100 ps (10 GHz). Table \ref{table_comp} presents a performance comparison with boolean and in-memory XOR compute macros.

The proposed X-pSRAM architecture not only supports memory storage but also enables in-memory Boolean XOR operations. While prior works such as \cite{eo_xor_dc1, eo_xor_dc3} demonstrate XOR functionality, they lack integrated memory, thereby continuing to suffer from the memory-wall bottleneck in data-intensive tasks. Conversely, \cite{optical_FF} presents an optical memory solution, but it does not incorporate computing capabilities. Electrical SRAM-based in-memory computing benefits from compact size due to mature CMOS technologies, yet it falls short in latency performance compared to photonic systems. In contrast, our XOR-augmented photonic memory, implemented using the GlobalFoundries 45SPCLO PDK, achieves a smaller layout footprint than existing photonic XOR or memory structures, offering improved scalability for next-generation photonic in-memory computing platforms, in addition to better latency performance compared to its electrical counterparts.

Due to inevitable local geometric mismatches during fabrication, the resonance wavelengths of microring resonators can deviate from their intended values. To compensate for these variations and ensure proper resonance with the incoming optical signal, integrated thermal tuning ports (i.e., VTH1, and VTH2, as shown in Fig. \ref{fig_xor}) are employed. These thermal heaters allow fine-tuning of the resonators within a half FSR, potentially requiring up to 7.2 mW of DC power per ring. However, the actual power needed depends on the degree of the wavelength mismatch. The inherent robustness and regenerative behavior of the latch-based structure, as discussed in \cite{pSRAM_design}, help tolerate moderate variations (\textpm 25 pm) in resonance wavelength of the latch microring. Additionally, careful layout practices—such as placing the rings in close physical proximity—can further minimize mismatches and reduce the need for significant thermal tuning. Moreover, thermal and supply voltage fluctuations can cause slight shifts in the resonance wavelength of the microring resonator. If these shifts exceed the acceptable detection or latching range, thermal calibration can be applied to restore proper functionality.


\subsection{Potential Applications}
Our proposed XOR-augmented photonic SRAM can serve as a core computing macro for various applications, including binary neural networks, where input data and weights are processed in an array-based multiply-accumulate manner \cite{xor_net, xnor_net, xnor_bnn}. As discussed earlier, wavelength-multiplexed signals can be summed in the electrical domain via a photodiode at the output, facilitating seamless integration with electrical subsystems. Additionally, by leveraging multi-FSR operation \cite{multi_FSR}, multiple input signals can be encoded at different free spectral ranges (FSRs), such as \si{\lambda_1}, \si{\lambda_1 + \Delta \lambda}, \si{\lambda_1 + 2\Delta \lambda}, and so on (where \si{\Delta \lambda} represents one FSR). This approach enables massively parallel computation by fixing weights while encoding different inputs at separate FSRs.  Furthermore, hyperdimensional computing relies on search operations across entire memory arrays \cite{hd_computing1, hd_computing2}, which fundamentally involve XOR computations. The proposed photonic in-memory XOR design can efficiently perform these searches with ultra-low energy consumption and high speed. Beyond neural networks and search operations, this photonic XOR mechanism can be applied to security applications, including encryption and decryption \cite{secure_XOR_CIM}, as well as copy and verification operations \cite{xor_verification}. Notably, due to the differential input configuration in our X-pSRAM, the same architecture can also execute XNOR operations by simply altering the polarity of input signals: setting logic-1 as \si{X = P_0}, \si{XB = P_X} and logic-0 as \si{X = P_X}, \si{XB = P_0}. Under these conditions, the output node \si{Z} produces active high XNOR.  

\section{Conclusion} \label{sec_conclusion}
In summary, we introduce and validate a novel XOR-augmented photonic SRAM (X-pSRAM) design that can perform in-situ Boolean XOR computation between the stored and input data. This in-situ structure is scalable to larger arrays through the use of wavelength-division multiplexing, enabling n-bit XOR computations in a single shot with massively parallel execution, thus enhancing computational throughput and efficiency. We have validated the functionality of the bitcell, including both memory and compute operations, using GlobalFoundries' 45SPLCO PDK. The CMOS-compatible, fabrication-friendly design of the X-pSRAM enables seamless integration with existing CMOS platforms and compatibility with electrical subsystems. This proposed solution offers a promising pathway for next-generation, data-intensive applications, including neural network acceleration, hyperdimensional computing, security, and verification tasks, paving the way for energy-efficient optical computing systems.

\section*{Acknowledgments}
This work is supported by the Defense Advanced Research Projects Agency (DARPA) under Grant No. HR001123S0024.

\bibliographystyle{IEEEtran}
\bibliography{ref}

\begin{thebibliography}{10}
\providecommand{\url}[1]{#1}
\csname url@samestyle\endcsname
\providecommand{\newblock}{\relax}
\providecommand{\bibinfo}[2]{#2}
\providecommand{\BIBentrySTDinterwordspacing}{\spaceskip=0pt\relax}
\providecommand{\BIBentryALTinterwordstretchfactor}{4}
\providecommand{\BIBentryALTinterwordspacing}{\spaceskip=\fontdimen2\font plus
\BIBentryALTinterwordstretchfactor\fontdimen3\font minus \fontdimen4\font\relax}
\providecommand{\BIBforeignlanguage}[2]{{%
\expandafter\ifx\csname l@#1\endcsname\relax
\typeout{** WARNING: IEEEtran.bst: No hyphenation pattern has been}%
\typeout{** loaded for the language `#1'. Using the pattern for}%
\typeout{** the default language instead.}%
\else
\language=\csname l@#1\endcsname
\fi
#2}}
\providecommand{\BIBdecl}{\relax}
\BIBdecl

\bibitem{memory_wall_bottleneck}
W.~A. Wulf \emph{et~al.}, ``Hitting the memory wall: Implications of the obvious,'' \emph{ACM SIGARCH computer architecture news}, vol.~23, no.~1, pp. 20--24, 1995.

\bibitem{imc_ref1}
A.~Sebastian \emph{et~al.}, ``Memory devices and applications for in-memory computing,'' \emph{Nature nanotechnology}, vol.~15, no.~7, pp. 529--544, 2020.

\bibitem{imc_ref2}
N.~Verma \emph{et~al.}, ``In-memory computing: Advances and prospects,'' \emph{IEEE Solid-State Circuits Magazine}, vol.~11, no.~3, pp. 43--55, 2019.

\bibitem{interconnect_prb1}
K.~Cho \emph{et~al.}, ``Sram write-and performance-assist cells for reducing interconnect resistance effects increased with technology scaling,'' \emph{IEEE Journal of Solid-State Circuits}, vol.~57, no.~4, pp. 1039--1048, 2022.

\bibitem{interconnect_prb2}
P.~Oldiges \emph{et~al.}, ``Chip power-frequency scaling in 10/7nm node,'' \emph{IEEE Access}, vol.~8, pp. 154\,329--154\,337, 2020.

\bibitem{optical_comp}
X.-Y. Xu \emph{et~al.}, ``Integrated photonic computing beyond the von neumann architecture,'' \emph{ACS Photonics}, vol.~10, no.~4, pp. 1027--1036, 2023.

\bibitem{optics_advantage1}
L.~M. Shaker \emph{et~al.}, ``Integrated photonics: bridging the gap between optics and electronics for enhancing information processing,'' \emph{Journal of Optics}, pp. 1--13, 2023.

\bibitem{optics_advantage2}
C.~G. Kibebe \emph{et~al.}, ``Harnessing optical advantages in computing: a review of current and future trends,'' \emph{Frontiers in Physics}, vol.~12, p. 1379051, 2024.

\bibitem{multiplex_optical_domain}
P.~J. Winzer, ``Modulation and multiplexing in optical communications,'' in \emph{Conference on Lasers and Electro-Optics}.\hskip 1em plus 0.5em minus 0.4em\relax Optica Publishing Group, 2009, p. CTuL3.

\bibitem{xnor_net}
M.~Rastegari \emph{et~al.}, ``Xnor-net: Imagenet classification using binary convolutional neural networks,'' in \emph{European conference on computer vision}.\hskip 1em plus 0.5em minus 0.4em\relax Springer, 2016, pp. 525--542.

\bibitem{xor_net}
S.~Zhu \emph{et~al.}, ``Xor-net: An efficient computation pipeline for binary neural network inference on edge devices,'' in \emph{2020 IEEE 26th international conference on parallel and distributed systems (ICPADS)}.\hskip 1em plus 0.5em minus 0.4em\relax IEEE, 2020, pp. 124--131.

\bibitem{xnor_bnn}
Z.~Wang \emph{et~al.}, ``Learning channel-wise interactions for binary convolutional neural networks,'' in \emph{Proceedings of the IEEE/CVF conference on computer vision and pattern recognition}, 2019, pp. 568--577.

\bibitem{secure_XOR_CIM}
S.~Huang \emph{et~al.}, ``Secure xor-cim engine: Compute-in-memory sram architecture with embedded xor encryption,'' \emph{IEEE Transactions on Very Large Scale Integration (VLSI) Systems}, vol.~29, no.~12, pp. 2027--2039, 2021.

\bibitem{xor_in_AI}
B.~Yan \emph{et~al.}, ``A 1.041-mb/mm 2 27.38-tops/w signed-int8 dynamic-logic-based adc-less sram compute-in-memory macro in 28nm with reconfigurable bitwise operation for ai and embedded applications,'' in \emph{2022 IEEE International Solid-State Circuits Conference (ISSCC)}, vol.~65.\hskip 1em plus 0.5em minus 0.4em\relax IEEE, 2022, pp. 188--190.

\bibitem{xnor_sram}
S.~Yin \emph{et~al.}, ``Xnor-sram: In-memory computing sram macro for binary/ternary deep neural networks,'' \emph{IEEE Journal of Solid-State Circuits}, vol.~55, no.~6, pp. 1733--1743, 2020.

\bibitem{xor_verification}
J.~Dreier \emph{et~al.}, ``Automated unbounded verification of stateful cryptographic protocols with exclusive or,'' in \emph{2018 IEEE 31st Computer Security Foundations Symposium (CSF)}.\hskip 1em plus 0.5em minus 0.4em\relax IEEE, 2018, pp. 359--373.

\bibitem{x_sram}
A.~Agrawal \emph{et~al.}, ``X-sram: Enabling in-memory boolean computations in cmos static random access memories,'' \emph{IEEE Transactions on Circuits and Systems I: Regular Papers}, vol.~65, no.~12, pp. 4219--4232, 2018.

\bibitem{xor_copy}
Z.~Lin \emph{et~al.}, ``In situ storing 8t sram-cim macro for full-array boolean logic and copy operations,'' \emph{IEEE Journal of Solid-State Circuits}, vol.~58, no.~5, pp. 1472--1486, 2022.

\bibitem{hd_computing1}
L.~Ge \emph{et~al.}, ``Classification using hyperdimensional computing: A review,'' \emph{IEEE Circuits and Systems Magazine}, vol.~20, no.~2, pp. 30--47, 2020.

\bibitem{hd_computing2}
E.~Hassan \emph{et~al.}, ``Hyper-dimensional computing challenges and opportunities for ai applications,'' \emph{IEEE Access}, vol.~10, pp. 97\,651--97\,664, 2021.

\bibitem{PhC_xor1}
H.~Wang \emph{et~al.}, ``All-optical and, xor, and not logic gates based on y-branch photonic crystal waveguide,'' \emph{Optical Engineering}, vol.~54, no.~7, pp. 077\,101--077\,101, 2015.

\bibitem{PhC_xor2}
R.~Rigi \emph{et~al.}, ``Configurable all-optical photonic crystal xor/and and xnor/nand logic gates,'' \emph{Optical and Quantum Electronics}, vol.~52, pp. 1--11, 2020.

\bibitem{PhC_xor3}
W.~Liu \emph{et~al.}, ``Design of ultra compact all-optical xor, xnor, nand and or gates using photonic crystal multi-mode interference waveguides,'' \emph{Optics \& Laser Technology}, vol.~50, pp. 55--64, 2013.

\bibitem{PhC_xor4}
C.~Husko \emph{et~al.}, ``Ultracompact all-optical xor logic gate in a slow-light silicon photonic crystal waveguide,'' \emph{Optics express}, vol.~19, no.~21, pp. 20\,681--20\,690, 2011.

\bibitem{eo_xor_dc1}
L.~Zhang \emph{et~al.}, ``Electro-optic directed logic circuit based on microring resonators for xor/xnor operations,'' \emph{Optics Express}, vol.~20, no.~11, pp. 11\,605--11\,614, 2012.

\bibitem{eo_xor_dc2}
L.~Zhang, R.~Ji \emph{et~al.}, ``Simultaneous implementation of xor and xnor operations using a directed logic circuit based on two microring resonators,'' \emph{Optics express}, vol.~19, no.~7, pp. 6524--6540, 2011.

\bibitem{eo_xor_dc3}
L.~Yang \emph{et~al.}, ``Xor and xnor operations at 12.5 gb/s using cascaded carrier-depletion microring resonators,'' \emph{Optics express}, vol.~22, no.~3, pp. 2996--3012, 2014.

\bibitem{mrm_ref1}
J.~Sun \emph{et~al.}, ``A 128 gb/s pam4 silicon microring modulator with integrated thermo-optic resonance tuning,'' \emph{Journal of Lightwave Technology}, vol.~37, no.~1, pp. 110--115, 2018.

\bibitem{mrm_ref2}
J.~Nijem \emph{et~al.}, ``High-q and high finesse silicon microring resonator,'' \emph{Optics Express}, vol.~32, no.~5, pp. 7896--7906, 2024.

\bibitem{pSRAM_design}
M.~A.-A. Kaiser \emph{et~al.}, ``Design of energy-efficient cross-coupled differential photonic-sram (psram) bitcell for high-speed on-chip photonic memory and compute systems,'' \emph{arXiv preprint arXiv:2503.19544}, 2025.

\bibitem{optical_FF}
S.~Pitris, C.~Vagionas, T.~Tekin, R.~Broeke, G.~Kanellos, and N.~Pleros, ``Wdm-enabled optical ram at 5 gb/s using a monolithic inp flip-flop chip,'' \emph{IEEE Photonics Journal}, vol.~8, no.~2, pp. 1--7, 2016.

\bibitem{multi_FSR}
X.~Xiao \emph{et~al.}, ``Wavelength-parallel photonic tensor core based on multi-fsr microring resonator crossbar array,'' in \emph{Optical Fiber Communication Conference}.\hskip 1em plus 0.5em minus 0.4em\relax Optica Publishing Group, 2023, pp. W3G--4.

\end{thebibliography}

\end{document}